\documentclass[superscriptaddress,prl,twocolumn,nofootinbib,showpacs,preprintnumbers]{revtex4}

\usepackage{graphicx,epsfig}

\def\bwt{\begin{widetext}}

\def\ewt{\end{widetext}}

\def\be{\begin{equation}}

\def\ee{\end{equation}}

\def\bea{\begin{eqnarray}}

\def\eea{\end{eqnarray}}

\def\bean{\begin{eqnarray*}}

\def\eean{\end{eqnarray*}}

\def\bary{\begin{array}}

\def\eary{\end{array}}

\def\bit{\begin{itemize}}

\def\eit{\end{itemize}}

\def\su5u1{SU(5) \times U(1)}

\def\fsu5u1{SU(5) \times U(1)'}

\def\so10{SO(10)}

\def\sq20{SO(10) \times SO(10)}

\usepackage[centertags]{amsmath}

\usepackage{amssymb}

\newcommand{\Z}{{\mathbb Z}}

\begin{document}

\title{Realistic Type IIB Supersymmetric Minkowski Flux Vacua}

\author{Ching-Ming Chen}

\affiliation{George P. and Cynthia W. Mitchell Institute for
Fundamental Physics, Texas A$\&$M University, College Station, TX
77843, USA }

\author{Tianjun Li}

\affiliation{George P. and Cynthia W. Mitchell Institute for
Fundamental Physics, Texas A$\&$M University, College Station, TX
77843, USA }

\affiliation{ Institute of Theoretical Physics, Chinese Academy of
Sciences, Beijing 100080, China}

\author{Yan Liu}

\affiliation{ Institute of Theoretical Physics, Chinese Academy of
Sciences, Beijing 100080, China}

\author{Dimitri V. Nanopoulos}

\affiliation{George P. and Cynthia W. Mitchell Institute for
Fundamental Physics, Texas A$\&$M University, College Station, TX
77843, USA }

\affiliation{Astroparticle Physics Group, Houston Advanced
Research Center (HARC), Mitchell Campus, Woodlands, TX 77381, USA}

\def\athens{Academy of Athens, Division of Natural Sciences, 
28 Panepistimiou Avenue, Athens 10679, Greece}

\affiliation{\athens}



\begin{abstract}

We show that there exist supersymmetric Minkowski vacua 
on Type IIB toroidal orientifold with general flux 
compactifications where the RR tadpole cancellation 
conditions can be relaxed elegantly. Then we present 
a realistic Pati-Salam like model. 
At the string scale, the gauge symmetry can be broken 
down to the Standard Model (SM) gauge symmetry, the 
gauge coupling unification can be achieved naturally, 
and all the extra chiral exotic particles can be decoupled 
so that we have the supersymmetric SMs with/without 
SM singlet(s) below the string scale. The observed 
SM fermion masses and mixings can also be obtained. 
In addition, the unified gauge coupling, the dilaton, 
the complex structure moduli, the real parts of 
the K\"ahler moduli and the sum of the imaginary parts 
of the K\"ahler moduli can be determined as functions 
of the four-dimensional dilaton and fluxes, and 
can be estimated as well.

\end{abstract}


\preprint{ACT-07-07, MIFP-07-30}

\maketitle


{\bf Introduction~--}~One of the great challenging and essential
problems in string phenomenology is the construction of
the realistic string vacua, which can give us the low energy
 supersymmetric Standard Models (SMs) without exotic particles,
and can stabilize the moduli fields. With renormalization group 
equation running, we can connect such constructions
to the low energy realistic particle physics  which will 
be tested at the upcoming Large Hadron Collider (LHC).  During the last 
a few years, the intersecting D-brane models on Type II orientifolds~\cite{JPEW}, 
where the chiral fermions arise from the intersections of D-branes in the
internal space~\cite{bdl} and the T-dual description in terms of
magnetized D-branes~\cite{bachas}, have been particularly
interesting~\cite{Blumenhagen:2005mu}.

On Type IIA orientifolds with intersecting D6-branes, many 
non-supersymmetric three-family Standard-like models and 
Grand Unified Theories (GUTs) were constructed in the 
beginning~\cite{Blumenhagen:2000wh}. However, there generically 
existed uncancelled Neveu-Schwarz-Neveu-Schwarz (NSNS) tadpoles
 and the gauge hierarchy problem. To solve these problems,
semi-realistic supersymmetric Standard-like and GUT models have
been constructed in Type IIA theory on the
$\mathbf{T^6/(\Z_2\times \Z_2)}$ orientifold~\cite{CSU,CLL} and
other backgrounds~\cite{ListSUSYOthers}. Interestingly,
only the Pati-Salam like models can give all the Yukawa couplings.
Without the flux background, Pati-Salam like models have been 
constructed systematically in Type IIA theory on the 
$\mathbf{T^6/(\Z_2\times \Z_2)}$  orientifold~\cite{CLL}. Although
we may explain the SM fermion masses and mixings in one 
model~\cite{Chen:2007px},  
the moduli fields have not been stabilized, and it is very 
difficult to decouple the chiral exotic particles.
To stabilize the moduli via supergravity fluxes, 
the flux models on Type II orientifolds have also been 
constructed~\cite{ListFlux,Chen:2006gd}. Especially, 
some flux models~\cite{Chen:2006gd} can explain the
SM fermion masses and mixings.
However, those models are in the AdS vacua and have
quite a few chiral exotic particles that are difficult 
to be decoupled.

In this paper, we consider the Type IIB toroidal
orientifold with the Ramond-Ramond (RR), NSNS, 
non-geometric and S-dual flux compactifications~\cite{Aldazabal:2006up}. 
We find that the RR tadpole 
cancellation conditions can be relaxed elegantly 
in the supersymmetric Minkowski vacua, and then
we may construct the realistic Pati-Salam like models~\cite{CLLN-L}. 
In this paper, we present a concrete simple model which 
is very interesting from the phenomenological point of view and 
might describe Nature. We emphasize that we do not fix the 
four-dimensional dilaton via flux potential. The point is that 
the fixed values for dilaton and K\"ahler moduli from flux 
compactifications are not consistent with those from the interesting
D-brane models and predict the wrong gauge
couplings at the string scale for the other models~\cite{CLLN-L}. 
This is a blessing in disguise from a cosmological 
point of view~\cite{Lahanas:2006xv}.

{\bf Type IIB Flux Compactifications~--}~We consider the Type IIB 
string theory compactified on a $\mathbf{T^6}$
orientifold where $\mathbf{T^{6}}$ is a six-torus
factorized as $\mathbf{T^{6}} = \mathbf{T^2} \times \mathbf{T^2}
\times \mathbf{T^2}$ whose complex coordinates are $z_i$, $i=1,\;
2,\; 3$ for the $i^{th}$ two-torus, respectively.
The orientifold projection is implemented by gauging the symmetry
$\Omega R$, where $\Omega$ is world-sheet parity, and $R$ is given
by
\begin{eqnarray}
R: (z_1,z_2,z_3) \to (-z_1, -z_2, -z_3)~.~\,    \label{orientifold}
\end{eqnarray}
Thus, the model contains 64 O3-planes. 
In order to cancel the negative RR charges from these
O3-planes, we introduce the magnetized  
D(3+2n)-branes which are filling up the 
four-dimensional Minkowski space-time and wrapping
2n-cycles on the compact manifold. Concretely, for one stack
of $N_a$ D-branes wrapped $m_a^i$ times on the $i^{th}$ 
two-torus $\mathbf{T^2_i}$, we turn on $n_a^i$ units of magnetic fluxes 
$F^i_a$ for the center of mass $U(1)_a$ gauge factor on $\mathbf{T^2_i}$, 
such that
\begin{eqnarray}
m_a^i \, \frac 1{2\pi}\, \int_{T^2_{\,i}} F_a^i \, = \, n_a^i ~,~\,
\label{monopole}
\end{eqnarray}
where $m_a^i$ can be half integer for tilted two-torus.
Then, the D9-, D7-, D5- and D3-branes contain 0, 1, 2 and 3 vanishing 
$m_a^i$s, respectively. Introducing for the $i^{th}$ two-torus 
the even homology classes $[{\bf 0}_i]$ and $[{\bf T}^2_i]$ for 
the point and two-torus, respectively, the vectors of the RR 
charges of the $a^{th}$ stack of D-branes and its image are
\begin{eqnarray}
 && [{ \Pi}_a]\, =\, \prod_{i=1}^3\, ( n_a^i [{\bf 0}_i] + m_a^i [{\bf T}^2_i] ), 
 \nonumber\\&&
[{\Pi}_a']\, =\, \prod_{i=1}^3\, ( n_a^i [{\bf 0}_i]- m_a^i [{\bf T}^2_i] )~,~
\label{homology class for D-branes}
\end{eqnarray}
respectively.
The ``intersection numbers'' in Type IIA language, which determine 
the chiral massless spectrum, are
\begin{eqnarray}
I_{ab}&=&[\Pi_a]\cdot[\Pi_b]=\prod_{i=1}^3(n_a^im_b^i-n_b^im_a^i)~.~
\label{intersections}
\end{eqnarray}
Moreover, for a stack of $N$ D(2n+3)-branes whose homology classes
on $\mathbf{T^{6}}$ is (not) invariant under $\Omega R$, we obtain 
a $USp(2N)$ ($U(N)$) gauge symmetry with three anti-symmetric 
(adjoint) chiral superfields due to the orbifold projection.  
The physical spectrum is presented in Table \ref{spectrum}.

\begin{table}[t]
\caption{General spectrum  for magnetized D-branes on the  Type IIB
${\mathbf{T^6}}$ orientifold. }
\renewcommand{\arraystretch}{1.25}
\begin{center}
\begin{tabular}{|c|c|}
\hline {\bf Sector} & {\bf Representation}
 \\
\hline\hline
$aa$   & $U(N_a)$ vector multiplet  \\
       & 3 adjoint multiplets  \\
\hline
$ab+ba$   & $I_{ab}$ $(N_a,{\overline{N}}_b)$ multiplets  \\
\hline
$ab'+b'a$ & $I_{ab'}$ $(N_a,N_b)$ multiplets \\
\hline $aa'+a'a$ &$\frac 12 (I_{aa'} -  I_{aO3})\;\;$
symmetric multiplets \\
          & $\frac 12 (I_{aa'} +  I_{aO3}) \;\;$ 
anti-symmetric multiplets \\
\hline
\end{tabular}
\end{center}
\label{spectrum}
\end{table}

The flux models  on Type IIB orientifolds with 
four-dimensional $N=1$ supersymmetry  are primarily
constrained by the  RR tadpole cancellation conditions that
will be given later, the four-dimensional $N=1$ supersymmetric
D-brane configurations, and the K-theory anomaly free conditions.
For the D-branes with world-volume magnetic field
$F_a^i={n_a^i}/({m_a^i\chi_i})$ where $\chi_i$ is the area of $\mathbf{T^2_i}$
in string units,  the condition for the four-dimensional 
$N=1$ supersymmetric D-brane configurations is 
\begin{eqnarray}
\sum_i \left(\tan^{-1} (F_a^i)^{-1} + 
{\theta (n_a^i)} \pi \right)=0 ~~~{\rm mod}~ 2\pi~,~\,
\end{eqnarray}
where ${\theta (n_a^i)}=1$ for $n_a^i < 0$ and  
${\theta (n_a^i)}=0$ for $n_a^i \geq 0$.
The K-theory anomaly free conditions are
\begin{eqnarray}
&&  \sum_a N_a m_a^1 m_a^2 m_a^3 =  \sum_a N_a m_a^1 n_a^2 n_a^3
= \sum_a N_a n_a^1 m_a^2 n_a^3 
\nonumber\\&&
= \sum_a N_a n_a^1 n_a^2 m_a^3
=0 ~~~{\rm mod}~ 2~.~\,
\end{eqnarray}
And the holomorphic gauge kinetic function for a generic stack of
D(2n+3)-branes  is given by~\cite{CLLN-L,Lust:2004cx}
\begin{eqnarray}
f_a &=& {1\over {\kappa_a}}\left(  n_a^1\,n_a^2\,n_a^3\,s-
n_a^1\,m_a^2\,m_a^3\,t_1 \right.\nonumber\\&& \left.
-n_a^2\,m_a^1\,m_a^3\,t_2 -n_a^3\,m_a^1\,m_a^2\,t_3\right)~,~\,
\label{EQ-GKF}
\end{eqnarray}
where $\kappa_a$ is equal to  1 and 2 for $U(n)$ and $USp(2n)$,
respectively.



We turn on the NSNS flux $h_0$, RR flux $e_i$,  
non-geometric fluxes  $b_{ii}$ and ${\bar b}_{ii}$, and the
S-dual fluxes $f_i$, $g_{ij}$ 
and $g_{ii}$~\cite{Aldazabal:2006up}. To avoid the subtleties,
these fluxes should be even integers due to the Dirac quantization.
For simplicity, we assume 
\begin{eqnarray}
&& e_i=e~,~~b_{ii}=\beta~,~~ {\bar b}_{ii}={\bar \beta}~,~~
\nonumber\\&&
f_i=f~,~ g_{ij}=-g_{ii}=g~,~\,
\end{eqnarray}
where $i\not= j$.
Then the constraint on fluxes from Bianchi indetities is
\begin{eqnarray}
f {\bar \beta} ~=~g \beta~.~\,
\end{eqnarray}
The RR tadpole cancellation conditions are
\begin{eqnarray}
&& \sum_a N_an_a^1 n_a^2 n_a^3 =16 ~,~
\nonumber\\&&
\sum_a N_a n_a^i m_a^j m_a^k = -{1\over 2} e {\bar \beta}~,~
\nonumber\\&&
N_{{\rm NS7}_i}=0~,~~N_{{\rm I7}_i}=0~,~\,
\end{eqnarray}
where $i\not= j \not= k \not= i$, and the $N_{{\rm NS7}_i}$
and $N_{{\rm I7}_i}$ denote the NS7 brane charge and
the other 7-brane charge, respectively~\cite{Aldazabal:2006up}.
Thus, if $e{\bar \beta} < 0$, the RR tadpole cancellation
conditions are relaxed elegantly because $-e{\bar \beta}/2$ only
needs to be even integer.
Moreover, we have 7 moduli fields in the supergravity
theory basis, the dilaton $s$, three K\"ahler moduli
$t_i$, and three complex structure moduli
$u_i$. With the above fluxes, we can assume 
\begin{eqnarray}
 t\equiv t_1+t_2+t_3~,~~~
u_1=u_2=u_3 \equiv u~.~\,
\end{eqnarray}
Then the superpotential becomes
\begin{eqnarray}
{\cal W}=3 i e u + i h_0 s - t \left(\beta u - 
i {\bar \beta} u^2 \right) - s t \left(f- igu\right).~\,
\end{eqnarray}
The K\"ahler potential for these moduli is
\begin{eqnarray}
{\cal K} = -{\rm ln}(s+{\bar s})-\sum_{i=1}^3 {\rm ln} (t_i + {\bar t}_i)
-\sum_{i=1}^3 {\rm ln} (u_i + {\bar u}_i)~.~\,
\end{eqnarray}
In addition, the supergravity scalar potential is
\begin{eqnarray}
V = e^{\cal K} \left({\cal K}^{i {\bar j}} D_i {\cal W} 
D_{\bar j} {\cal W} -3|{\cal W}|^2 \right)~,~\,
\end{eqnarray}
where ${\cal K}^{i {\bar j}}$ is the inverse metric of 
${\cal K}_{i {\bar j}}\equiv \partial_i \partial_{\bar j} {\cal K}$,
 $D_i {\cal W} =\partial_i {\cal W} + (\partial_i {\cal K}) {\cal W}$,
and $\partial_i = \partial_{\phi_i}$ where $\phi_i$ can be $s$, $t_i$,
and $u_i$. Thus, for the supersymmetric Minkowski vacua, we have
\begin{eqnarray}
{\cal W}={\partial_s  {\cal W}} ={\partial_t  {\cal W}} 
={\partial_u  {\cal W}} =0 ~.~\,
\end{eqnarray}
From ${\partial_s  {\cal W}}={\partial_t  {\cal W}}=0$, we obtain 
\begin{eqnarray}
t= {{ih_0}\over {f-igu}}~,~~~ s=-{{\beta}\over f} u~,~\,
\label{Moduli-Relations}
\end{eqnarray}
 then the superpotential turns out
\begin{eqnarray}
{\cal W} = \left( 3e -{{h_0 \beta}\over f}\right) i u~.~\,
\end{eqnarray}
Therefore, to satisfy ${\cal W}={\partial_u  {\cal W}} =0$, we obtain
\begin{eqnarray}
3 e f = \beta h_0~.~\,
\end{eqnarray}
Because ${\rm Re} s > 0$,  ${\rm Re} t_i > 0$ and ${\rm Re} u_i > 0$,
we require 
\begin{eqnarray}
{{h_0}\over g} < 0~,~~~{{\beta}\over f} < 0~.~\,
\end{eqnarray}



{\bf Model~--}~Choosing $e {\bar \beta}=-12$, we present the 
D-brane configurations and intersection numbers 
in Table~\ref{MI-Numbers}, and the resulting spectrum in
Table~\ref{Spectrum-I}. The anomalies from three global $U(1)$s of 
$U(4)_C$, $U(2)_L$ and $U(2)_R$ 
are cancelled by the Green-Schwarz mechanism, and the gauge fields of 
these $U(1)$s obtain masses via the linear $B\wedge F$ couplings. So, the
effective gauge symmetry is $SU(4)_C\times SU(2)_L\times SU(2)_R$.
In order to break the gauge symmetry, on the first two-torus, 
we split the $a$ stack of
D-branes into $a_1$ and $a_2$ stacks with 3 and 1 D-branes,
respectively, and split the $c$ stack of D-branes into $c_1$ and
$c_2$ stacks with 1 D-brane for each one.
Then, the gauge symmetry is further broken down to 
$ SU(3)_C\times SU(2)_L\times U(1)_{I_{3R}}\times U(1)_{B-L}$.
We can break the $U(1)_{I_{3R}}\times U(1)_{B-L}$ gauge
symmetry down to the $U(1)_Y$ gauge symmetry by giving
vacuum expectation values
(VEVs) to the vector-like particles with quantum numbers
$({\bf { 1}, 1, 1/2, -1})$ and $({\bf { 1},
1, -1/2, 1})$ under $SU(3)_C\times SU(2)_L\times U(1)_{I_{3R}} \times
U(1)_{B-L} $ from $a_2 c_1'$ D-brane intersections. Similar to the
discussions in Ref.~\cite{Chen:2007px}, we can explain 
the SM fermion masses and mixings via 
the Higgs fields $H_u^i$, $H_u'$, $H_d^i$ and
$H_d'$ because all the SM fermions and Higgs fields arise 
from the intersections on the first torus. 
To decouple the chiral exotic particles, we assume that 
the $T_R^i$ and $S_R^i$ obtain VEVs
at about the string scale, and their VEVs satisfy the 
D-flatness $U(1)_R$. The chiral exotic particles can obtain
masses via the following superpotential
\begin{eqnarray}
W \supset {1\over {M_{\rm St}}} S_R^i S_R^j
 T_R^k T_R^l + T_R^i X^j X^k~,~\,
\end{eqnarray}
where $M_{\rm St}$ is the string scale, and
we neglect the ${\cal{O}}(1)$ coefficients in this paper. 
In addition, the vector-like particles $S_L^i$ and
$\overline{S}_L^i$ in the anti-symmetric 
representation of $SU(2)_L$  can obtain the VEVs close to 
the string scale while keeping the D-flatness $U(1)_L$. Thus, we can
decouple all the Higgs bidoublets close to the string scale  
except one pair of the linear combinations of the Higgs doublets 
for the electroweak symmetry breaking at the low
energy by fine-tuning the following superpotential
\begin{eqnarray}
W & \supset & \Phi_i ( \overline{S}_L^j \Phi' + S_R^j \overline{\Phi}')
+ \overline{\Phi}_i ( T_R^j \Phi' + S_L^j \overline{\Phi}')
\nonumber\\&&
+{1\over {M_{\rm St}}} \left( \overline{S}_L^i S_R^j \Phi_k \Phi_l +
S_L^i T_R^j \overline{\Phi}_k \overline{\Phi}_l 
\right.\nonumber\\&&\left.
+\overline{S}_L^i T_R^j \Phi' \Phi' 
+ S_L^i S_R^j \overline{\Phi}' \overline{\Phi}' \right)
~.~\,
\end{eqnarray}
In short, below the string scale, we have the supersymmetric
SMs which may have zero, one or a few SM singlets from $S_L^i$, 
$\overline{S}_L^i$, and/or $S_R^i$. And then the
low bound on the lightest CP-even Higgs boson mass in the
minimal supersymmetric SM can
be relaxed if we have the SM singlet(s) at low energy~\cite{Li:2006xb}.

\begin{table}[htb]
\footnotesize
\renewcommand{\arraystretch}{1.0}
\caption{D-brane configurations and intersection numbers.}
\label{MI-Numbers}
\begin{center}
\begin{tabular}{|c||c|c||c|c|c|c|c|c|c|}
\hline
 & \multicolumn{9}{c|}{$U(4)_C\times U(2)_L\times U(2)_R\times USp(10)$}\\
\hline \hline  & $N$ & $(n^i,m^i)$ & $n_{S}$& $n_{A}$ & $b$ & $b'$ & $c$ & $c'$& $O3$ \\

\hline\hline

$a$ & 4 & $( 1, 0) \times ( 1,-1/2) \times ( 1, 1)$ & 0 & 0 & 3 & 0(3) & -3 &
0(3) & 0(1)
\\ \hline

$b$ & 2 & $( 1,-3) \times ( 1, 1/2) \times ( 1, 0)$ & 0 & 0(6) & - & - & 0(6) &
0(1) & 0(3)   \\ \hline

$c$ & 2 & $( 1, 3) \times ( 1, 1/2) \times ( 0,-1)$ & -6 & 6 & - & - & - & - & 3 \\

\hline \hline

$O3$ & 5 & $( 1, 0) \times ( 2, 0) \times ( 1, 0)$ & - & - &
\multicolumn{5}{|c|}{$6\chi_1 = \chi_2= 2\chi_3=2$}
 \\ \hline

\end{tabular}

\end{center}

\end{table}

\begin{table}[htb]
\footnotesize
\renewcommand{\arraystretch}{1.0}
\caption{The chiral and vector-like superfields, and their quantum
numbers under the gauge symmetry $SU(4)_C\times SU(2)_L\times
SU(2)_R \times USp(10)$.} \label{Spectrum-I}
\begin{center}
\begin{tabular}{|c||c||c|c|c||c|c|c|}\hline
 & Quantum Number
& $Q_4$ & $Q_{2L}$ & $Q_{2R}$  & Field \\
\hline\hline
$ab$ & $3 \times (4,\bar{2},1,1)$ & 1 & -1 & 0  & $F_L(Q_L, L_L)$\\
$ac$ & $3\times (\bar{4},1,2,1)$ & -1 & 0 & $1$   & $F_R(Q_R, L_R)$\\
$c_{S}$ & $6\times(1,1,\bar{3},1,1)$ & 0 & 0 & -2   & $T_R^i$  \\
$c_{A}$ & $6\times(1,1,1,1,1)$ & 0 & 0 & 2   & $S_R^i$ \\
$cO3$ & $3\times(1,1,2,10)$ & 0 & 0 & 1   &  $X^i$ \\
\hline\hline
$ac'$ & $3 \times (4,1,2,1)$ & 1 & 0 & 1  &  \\
& $3 \times (\bar{4}, 1, \bar{2},1)$ & -1 & 0 & -1 & \\
\hline
$bc$ & $6 \times (1,2,\overline{2},1)$ & 0 & 1 & -1   & 
$\Phi_i$($H_u^i$, $H_d^i$)\\
& $6 \times (1,\overline{2},2,1)$ & 0 & -1 & 1   & $\overline{\Phi}_i$ \\
\hline
$bc'$ & $1 \times (1,2, 2,1)$ & 0 & 1 & 1   & $\Phi'$($H'_u$, $H'_d$)\\
& $1 \times (1,\overline{2}, \overline{2},1)$ & 0 & -1 & -1   & 
$\overline{\Phi}'$ \\
\hline
$bb'$ 
 & $6 \times (1, 1, 1,1)$ & 0 & 2 & 0   & $S_L^i$ \\
 & $6 \times (1, \overline{1}, 1,1)$ & 0 & -2 & 0   & $\overline{S}_L^i$ \\
\hline
\end{tabular}
\end{center}
\end{table}

Next, we consider the gauge coupling unification and moduli stabilization.
The real parts of the dilaton and K\"ahler moduli in our model are~\cite{CLLN-L}
\begin{eqnarray}
&&  {\rm Re} s = {{{\sqrt 6} e^{-\phi_4}}\over {4 \pi}}~,~
{\rm Re} t_1 = {{{\sqrt 6} e^{-\phi_4}}\over {2 \pi}}~,~
\nonumber\\&&
{\rm Re} t_2 ={{{\sqrt 6} e^{-\phi_4}}\over {12 \pi}}~,~
{\rm Re} t_3 = {{{\sqrt 6} e^{-\phi_4}}\over {6 \pi}}~,~\,
\label{Moduli}
\end{eqnarray}
where $\phi_4$ is the four-dimensional dilaton. From Eq. (\ref{EQ-GKF}),
 we obtain that the SM gauge couplings are unified at the string
scale as follows
\begin{eqnarray}
g_{SU(3)_C}^{-2} = g_{SU(2)_L}^{-2} ={3\over 5} g_{U(1)_Y}^{-2}
={{{\sqrt 6} e^{-\phi_4}}\over {2 \pi}}~.~\,
\end{eqnarray}
Using the unified gauge coupling $g^{2} \simeq 0.513$
in supersymmetric SMs, we get
\begin{eqnarray}
 \phi_4 \simeq - 1.61~.~\,
\end{eqnarray}

For moduli stabilization, we first obtain $t$
from Eqs. (\ref{Moduli-Relations}) and (\ref{Moduli})
\begin{eqnarray}
{\rm Re} t = {{3{\sqrt 6} e^{-\phi_4}}\over {4 \pi}}
~,~ {\rm Im} t= \pm {\sqrt {{{3 \beta h_0}\over {fg}}- 
 {{27 e^{-2\phi_4}}\over {8 \pi^2}}}}  ~.~\,
\end{eqnarray}
Thus, we have
\begin{eqnarray}
&& {\rm Im} s = -{1\over 3} {\rm Im} t +{{\beta}\over g}~,~
\nonumber\\&&
{\rm Re} u = - {{{\sqrt 6} f e^{-\phi_4}}\over {4 \pi \beta}}~,~~
{\rm Im} u = {f\over {3\beta}} {\rm Im} t -{{f}\over g}~.~
\end{eqnarray}
Let us present a set of possible 
solutions to the fluxes 
\begin{eqnarray}
&& h_0=-18 \eta~,~~e= 6 \eta~,~~ \beta= 2 \eta'~,~
\nonumber\\&&
{\bar \beta}=- 2 \eta~,~~ f=-2 \eta'~,~~ g=2 \eta~,~
\end{eqnarray}
where $\eta = \pm 1$ and $\eta'=\pm 1$.
Choosing $\phi_4=-1.61$, $\eta =\eta'=1$, we obtain
the numerical values for the moduli fields 
\begin{eqnarray}
&& {\rm Re} s={\rm Re} u = 0.975~,~~~{\rm Re} t_1=1.95~,~
\nonumber\\&&
{\rm Re} t_2=0.325~,~~~{\rm Re} t_3=0.650~,~
\nonumber\\&&
\sum_{i=1}^3 {\rm Im} t_i = \pm 4.30~,~~~
{\rm Im} s = {\rm Im} u = \mp 1.43 +1 ~.~\,
\end{eqnarray}





{\bf Conclusions~--}~We showed that the RR tadpole cancellation 
conditions can be relaxed elegantly in the supersymmetric 
Minkowski vacua on the Type IIB toroidal orientifold with 
general flux compactifications. And we presented
a realistic Pati-Salam like model in details. In this
model, we can break 
the gauge symmetry down to the SM gauge symmetry,
realize the gauge coupling unification,
and decouple  all the extra chiral exotic particles
around the string scale. We can also generate the
observed SM fermion masses and mixings.
Futhermore, the unified gauge coupling, the dilaton, 
the complex structure moduli, the real parts of 
the K\"ahler moduli and the sum of the imaginary parts 
of the K\"ahler moduli can be determined as functions 
of the four-dimensional dilaton and fluxes,
and can also be estimated.

{\bf Acknowledgments~--}~This research was supported in part by
the Mitchell-Heep Chair in High Energy Physics (CMC), by the
Cambridge-Mitchell Collaboration in Theoretical Cosmology (TL),
and by the DOE grant DE-FG03-95-Er-40917 (DVN).


\end{document}